\documentclass[aps,prb,superscriptaddress,twocolumn]{revtex4}
\usepackage{graphicx}
\usepackage{epstopdf}
\usepackage{bm}
\usepackage{amssymb}
\usepackage{color}

\begin{document}

\title{
Chiral magnetism and helimagnons in a pyrochlore antiferromagnet}
\author{Eunsong Choi}
\affiliation{Department of Physics, University of Wisconsin, Madison, Wisconsin 53706, USA}

\author{Gia-Wei Chern}
\affiliation{Department of Physics, University of Wisconsin, Madison, Wisconsin 53706, USA}
\affiliation{Theoretical Division, Los Alamos National Laboratory, Los Alamos, New Mexico 87545, USA}

\author{Natalia B. Perkins}
\affiliation{Department of Physics, University of Wisconsin, Madison, Wisconsin 53706, USA}

\date{\today}

\begin{abstract}
Recent neutron scattering measurements on the spinel CdCr$_2$O$_4$ revealed a rare example of helical magnetic order in
geometrically frustrated pyrochlore antiferromagnet. The spin spiral characterized by an incommensurate
wavevector $\mathbf Q = 2\pi (0, \delta, 1)$ with $\delta \approx 0.09$ is accompanied by a tetragonal distortion.
Here we conduct a systematic study on the magnetic ground state resulting from the interplay between
the Dzyaloshinskii-Moriya interaction and further neighbor exchange couplings, two of the most important
mechanisms for stabilizing incommensurate spin orders. We compute the low-energy spin-wave spectrum based on a microscopic
spin Hamiltonian and find a dispersion relation characteristic of the helimagnons.
By numerically integrating the Landau-Lifshitz-Gilbert equation with realistic model parameters,
an overall agreement between experiment and the numerical spectrum, lending further support to the view that
a softened optical phonon triggers the magnetic transition and endows the lattice a chirality.
\end{abstract}

\maketitle

\section{Introduction}

Magnets with geometrical frustration provide a fertile ground for studying complex spin orders
and unconventional magnetic phases.\cite{moessner06} The defining feature of strong geometrical frustration
is the occurrence of extensively degenerate ground states. This usually arises when the arrangement of spins
on a lattice precludes satisfying all interactions at the same time. A great deal of attention has been
devoted to the so-called pyrochlore magnets in which the magnetic ions form a three-dimensional connected network of
corner-sharing tetrahedra.\cite{moessner98} With magnetic interactions restricted to only nearest neighbors,
classical spins on the pyrochlore lattice remain disordered even at temperatures well below the Curie-Weiss
energy scale. Magnetic fluctuations in this liquid-like phase are subject to strong local constraints that
maintain vanishing total spin on each tetrahedron. This disordered yet highly correlated phase is
often called a cooperative paramagnet. Because of the huge degeneracy, the magnet is sensitive to
nominally small perturbations. The low-temperature ordering thus depends subtly on the residual perturbations
present in the system.

The recent discovery of a novel chiral spin structure in spinel CdCr$_2$O$_4$
has generated much interest both theoretically and experimentally. \cite{chung05,matsuda07}
The magnetic Cr ions in this compound have spin $S=3/2$ and no
orbital degrees of freedom, providing a very good realization of Heisenberg antiferromagnet on the pyrochlore lattice.
Despite a rather high Curie-Weiss temperature $|\Theta_{\rm CW}| \approx 70$ K, a spiral magnetic order sets in only
at $T_N = 7.8$ K, indicating a high degree of frustration. The discontinuous magnetic transition at $T_N$
is accompanied by a structural distortion which lowers the crystal symmetry from cubic to tetragonal ($a=b < c$) with
an elongated unit cell: $(c-a)/c \approx 5\times 10^{-3}$.
Polarized neutron-scattering experiment reveals an incommensurate magnetic order in which coplanar spins rotate about
either $a$ or $b$ axes with a period of roughly ten lattice constants; the ordering wavevector $\mathbf Q = 2\pi (0, \delta, 1)$
with $\delta \approx 0.09$. It is worth noting that a very similar coplanar helical order has also been reported in the weak itinerant
antiferromagnet YMn$_2$~\cite{ballou87,cywinski91}.

Spin-lattice coupling has been shown to play an important role in relieving magnetic frustrations
in several chromium spinels. \cite{lee00,ueda06,matsuda07_Hg,rudolf07}
In particular, there are strong experimental evidences indicating that the magnetic transition in CdCr$_2$O$_4$ is
mainly driven by spin-lattice coupling. These include anomalies of the elastic constants, \cite{bhattacharjee11,zherlitsyn10}
splitting of optical phonons, \cite{valdes08,kant09} and energy shift of zone-boundary phonons \cite{kim11}
at the magnetic ordering temperature. The spin-lattice or magnetoelastic coupling arises from the dependence
of exchange on the ion displacements $\delta r_{ij}$ of spins:
$(\partial J/\partial r) (\mathbf S_i\cdot\mathbf S_j)\, \delta r_{ij}$. \cite{tcher02,chern09}
As a result, the lattice distortion creates disparities between nearest-neighbor exchange constants
and paves the way for magnetic ordering. However, spin-lattice coupling usually favors collinear magnetic orders.
Indeed, assuming a simple elastic energy cost $k(\delta r_{ij})^2/2$ and integrating out the displacements $\delta r_{ij}$
in the above model  gives rise to a biquadratic spin interaction: $-(\mathbf S_i\cdot\mathbf S_j)^2$, which is minimized by collinear spins.
While the magnetic frustration is relieved by the distortion, the appearance of helical spin order in CdCr$_2$O$_4$ thus comes
from other mechanisms.

Magnetic spirals resulting from the competition between nearest neighbor
and next nearest neighbor interactions are a common theme in frustrated spin systems. \cite{yoshimori59,adam80}
However, this is not the case for pyrochlore lattice.
Detailed mean-field and numerical investigations show that inclusion of further-neighbor
interactions partially relieves the frustration and sometimes selects rather complicated multiple-$\mathbf q$
spin orders, but fails to stabilize magnetic spirals with a definite handedness. \cite{reimers91,chern08}
Another mechanism for the formation of spirals is the Dzyaloshinskii-Moriya (DM)
interaction which originates from the relativistic spin-orbit coupling. \cite{dzyaloshinskii64,moriya60}
Although symmetry considerations allow the presence of DM terms in the pyrochlore lattice, \cite{elhajal05}
the ground states selected by the DM interaction are non-chiral magnetic orders with
wavevector $\mathbf q=0$. \cite{elhajal05,canals08,chern10}

The origin of the spiral order as well as the lattice distortion in CdCr$_2$O$_4$ has been consistently
explained in Ref.~\onlinecite{chern06}. In this scenario, the magnetic transition is triggered by the softening
of $\mathbf q = 0$ optical phonons with odd parity. The two types of tetrahedra with different orientations
are flattened along the $a$ and $b$ directions, respectively, resulting in an overall tetragonal elongation
along the $c$ axis. The lattice distortion relieves the magnetic frustration by creating
disparities among the nearest-neighbor exchange constants and stabilizes a collinear N\'eel order with
wavevector $\mathbf Q_0 = 2\pi(0,0,1)$.
Moreover, the broken parity due to the staggered distortion
gives rise to a {\em chiral} pyrochlore structure. The collinear spins are twisted into a long-period
spiral with $\mathbf Q = 2\pi(0,\delta, 1)$ as the structural chirality is transferred to magnetic order
by relativistic spin-orbit interaction.\cite{chern06,chern09}
The broken parity in CdCr$_2$O$_4$ has also been confirmed by recent optical measurements. \cite{kant09}

In this paper we undertake a systematic study of the general helical order resulting from the interplay between
DM interaction and further-neighbor couplings in pyrochlore antiferromagnets. We also employ various numerical methods
to investigate the spin-wave excitations in the spiral magnetic orders.
We first revisit the continuum theory which is valid in the $J \to \infty$ limit and show that the resultant
Hamiltonian possesses a continuous symmetry similar to that in smectic liquid crystals. \cite{deGennes93,chaikin,radzihovsky11}
In particular, despite the broken symmetry of the helical order is described by a O(3) order parameter,
the three associated Goldstone modes reduce to a single smectic-like phonon mode which emerges on scales
longer than the helical pitch. This low-energy gapless excitations, dubbed helimagnons in Ref.~\onlinecite{belitz06},
exhibit a highly anisotropic dispersion relation.

The emergent continuous symmetry of the effective theory is broken by the anisotropic crystal fields
at the microscopic level. We first adopt the spherical approximation to investigate the resultant magnetic ground states.
Three typical helical orders with wavevectors $(0, 0, 1-\delta)$, $(0, \delta, 1)$ and $(\delta, \delta, 1)$
are stabilized depending on the relative strengths of DM interaction $D$ and
third neighbor interaction $J_3$. We then focus on the $(0, \delta, 1)$ spiral,
which is relevant to the case of CdCr$_2$O$_4$, and consider perturbation corrections to the coplanar spiral
solution obtained from the continuum theory. We show that the first-order corrections do not shift the spiral pitch
but introduces a small out-of-plane spin component.

Next we study the magnetic excitations in the spiral magnetic ground state.
The incommensurate nature of the helical order presents a challenging task for the calculation
of spin-wave spectrum. To circumvent this problem we employ the conventional Holstein-Primakoff method
with a large commensurate unit cell. We also perform dynamics simulations based on the
Landau-Lifshitz-Gilbert (LLG) equations. \cite{mochizuki10}
The LLG simulations with a large damping also provide an efficient
approach to obtain the accurate spin structures in the equilibrium state. The magnon spectrum obtained from
our dynamics simulation exhibits a gapless mode at the helical wavevector $\mathbf Q = 2\pi(0, \delta, 1)$
reminiscent of the smectic-like phonon mode or helimagnons mentioned above. \cite{radzihovsky11,belitz06}
An overall agreement is obtained between the numerical calculations and the experimental data on CdCr$_2$O$_4$.

The remainder of this paper is organized as follows. In Sec.~\ref{sec:helical} we consider the helical magnetic order
in pyrochlore antiferromagnet. After presenting a minimal model relevant for CdCr$_2$O$_4$, we first review the
continuum theory of the magnetic spirals on pyrochlore.
We then demonstrate the existence of a continuous symmetry similar to that in smectic liquid crystals.
We obtain a phase diagram of the helical order based on the spherical approximation.
Sec.~III is devoted to a discussion of the magnon excitations of pyrochlore magnet.
A systematic analysis of the low-energy acoustic mode indicates a highly anisotropic
dispersion characteristic of that of helimagnons.
We then perform dynamics simulation to obtain the spin-wave spectrum of the helical spin order
with wavevector $\mathbf Q = 2\pi (0, \delta, 1)$ and compare the results with available experimental data.
We conclude in Sec.~\ref{sec:conclusion} with a summary and discussion of our results.

\begin{figure}
\centering{}\includegraphics[width=1.0\columnwidth]{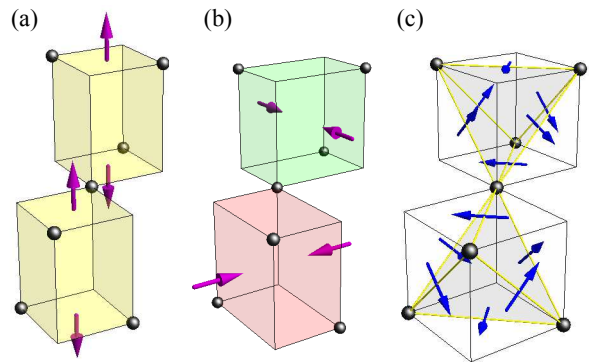}
\caption{(Color online) Schematic picture of (a) even and (b) odd tetragonal distortions with $E_g$ and $E_u$ symmetries,
respectively. Both types of distortion preserve the lattice translational symmetry
and could result in an overall elongation along the $c$ axis. Panel (c) shows
the DM vectors $\mathbf D_{ij}$ on the six inequivalent NN bonds of the pyrochlore lattice.}
\label{fig:distortion}
\end{figure}

\section{Helical magnetic order}
\label{sec:helical}

Our starting point is a classical Heisenberg model on the pyrochlore lattice described by the Hamiltonian
\begin{eqnarray}
	\label{eq:H_model}
	\mathcal{H} &=& \sum_{\langle ij \rangle} \left(J + K_{ij}\right)\,{\bf S}_{i}\cdot{\bf S}_{j}
	+ \sum_{\langle ij \rangle} {\bf D}_{ij}\cdot\left({\bf S}_{i}\times{\bf S}_{j}\right) \nonumber \\
	&& +\,\, J_2 \sum_{\langle\langle ij \rangle\rangle} \mathbf S_i\cdot\mathbf S_j	
	+ J_3 \sum_{\langle\langle\langle ij \rangle\rangle\rangle} \mathbf S_i\cdot\mathbf S_j,
\end{eqnarray}
The dominant term is the nearest-neighbor (NN) exchange interaction with an antiferromagnetic sign $J > 0$.
Due to the special geometry of the pyrochlore lattice, this term can be recast to
$(J/2)\sum_{\mathbf r} \left|\mathbf M(\mathbf r)\right|^2$,
where $\mathbf M$ denotes the vector sum of spins on a tetrahedron at $\mathbf r$.
Minimization of this term requires the vanishing of total magnetic moment on every tetrahedron but
leaves an extensively degenerate manifold.

The exchange anisotropy $K_{ij}$ results from a distorted lattice through magnetoelastic coupling.
The structural distortion in CdCr$_2$O$_4$ reduces the crystal symmetry from cubic to tetragonal
while preserving the lattice translational symmetry. \cite{chung05} Here we consider $\mathbf q = 0$ lattice distortions
with even and odd parities; see Fig.~\ref{fig:distortion}(a) and~(b). The dominant $E_u$ component
results in a staggered distortion: tetrahedra of types I and II are flattened along the $a$ and $b$ directions,
respectively; the lattice is elongated overall, $a = b < c$. The resulting space group, $I4_122$, lacks the inversion symmetry.
Fig.~\ref{fig:collinear} shows the exchange anisotropy of the six NN bonds in the unit cell.
As pointed out in Ref.~\onlinecite{tcher02}, this tetragonal distortion completely relieves the magnetic frustration
and stabilizes a collinear N\'eel order with wavevector $\mathbf Q_0 = 2\pi(0, 0, 1)$ shown in Fig.~\ref{fig:collinear}.

The DM term $\mathbf D_{ij}\cdot(\mathbf S_i\times\mathbf S_j)$ originates from the relativistic spin-orbit
interaction $\lambda (\mathbf L\cdot\mathbf S)$.
This term is allowed on the ideal pyrochlore lattice, where the bonds are not centrosymmetric as
required by the so-called Moriya rules. \cite{moriya60} Moreover, the high symmetry of the pyrochlore lattice
completely determines the orientations of the DM vectors [see Fig.~\ref{fig:distortion}(c)] up to a multiplicative factor.
Finally, we also include second and third exchanges couplings $J_2$ and $J_3$, respectively, in the model
Hamiltonian. First principle {\it ab initio} calculations indicate that these further-neighbor exchanges
are important for the modeling of spinel CdCr$_2$O$_4$. \cite{chern06,yaresko08}
In the limit dominated by a large NN exchange, the ground-state properties actually depend
only on the difference between $J_2$ and $J_3$. The relative shift in energy for any pair of ground states
due to a small $J_3$ is identical to the effect of a $J_2$ of the same magnitude with an opposite sign. \cite{chern08}

\subsection{Gradient approximation and emergent continuous symmetry}
\label{sec:continuum}

\begin{figure}
\centering{}\includegraphics[width=0.75\columnwidth]{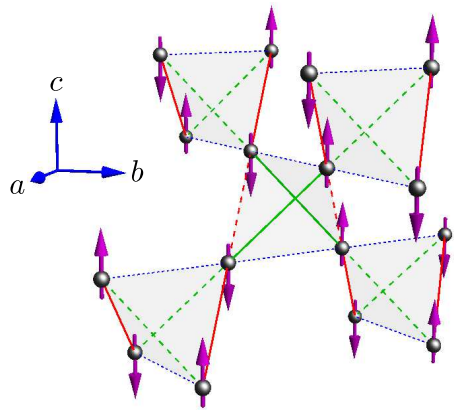}
\caption{(Color online) A N\'eel order on the pyrochlore lattice with wavevector $\mathbf q = 2\pi(0,0,1)$.
This collinear spin order is stabilized by a staggered distortion with $E_u$ symmetry.
Tetrahedra of type I and II are flattened along the $a$
and $b$ directions, respectively. The exchange constants on various NN bonds are
$J + K_u - K_g$ (solid), $J - 2K_u - K_g$ (dashed), and $J + K_u + 2K_g$ (dotted).} 
\label{fig:collinear}
\end{figure}

\begin{figure}
\centering{}\includegraphics[width=1\columnwidth]{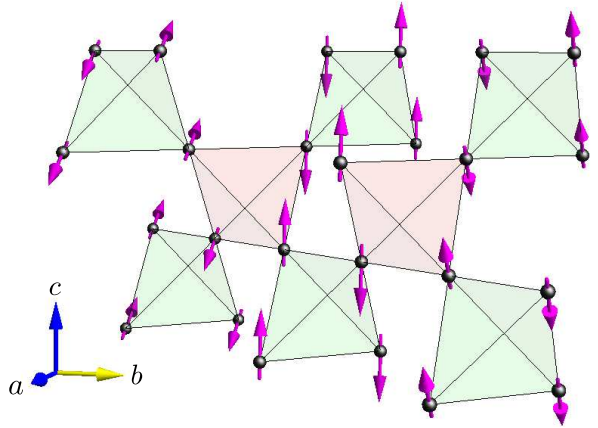}
\caption{(Color online) Coplanar helical order with spins lying in the $ac$ plane and rotating about the $b$ axis.
The magnetic order corresponds to the solution $\hat\mathbf n_1 = (\cos\theta(y), 0, \sin\theta(y))$,
with $\theta(y) = 2\pi\delta y + \mbox{const}$. It produces a Bragg peak $\mathbf Q = 2\pi(0, \delta, 1)$ in
neutron scattering, consistent with the experimental measurement on CdCr$_2$O$_4$.
\label{fig:spiral}}
\end{figure}

We first briefly review the continuum theory of helical orders derived from the $\mathbf Q_0 = 2\pi(0,0,1)$ N\'eel order.
The effective energy functional is obtained based on a gradient approximation of the staggered magnetizations.\cite{chern06,chern09}
The antiferromagnetic order on the pyrochlore lattice is characterized by three order parameters:
\begin{eqnarray}
	\mathbf L_1 &=& (\mathbf S_0 + \mathbf S_1 - \mathbf S_2 - \mathbf S_3)/4, \nonumber \\
	\mathbf L_2 &=& (\mathbf S_0 - \mathbf S_1 + \mathbf S_2 - \mathbf S_3)/4, \\
	\mathbf L_3 &=& (\mathbf S_0 - \mathbf S_1 - \mathbf S_2 + \mathbf S_3)/4, \nonumber
\end{eqnarray}
where $\mathbf S_\mu$ denotes magnetic moments at $\mu$th sublattice. \cite{chern06}
For example, the collinear N\'eel state shown in Fig.~\ref{fig:collinear} is
described by order parameters: $\mathbf M = 0$, $\mathbf L_2 = \mathbf L_3 = 0$,
and $\mathbf L_1 = {\hat {\mathbf n}}_1\,e^{i\mathbf Q_0\cdot\mathbf r}$, where ${\hat\mathbf n}_1$ is an arbitrary
unit vector.  Assuming a slow variation of staggered magnetizations in the low-period spiral, we parameterize the
order parameters as: $\mathbf L_i(\mathbf r) = e^{i\mathbf q\cdot\mathbf r}\phi_i(\mathbf r)\,\hat\mathbf n_i(\mathbf r)$, where $\mathbf n_i(\mathbf r)$ and $\phi_i(r)$ are the directions and magnitudes of the staggered magnetizations.
The hard constraints $|\mathbf S_i| = S$ require that three unit vectors $\hat\mathbf n_i$ are orthogonal to each other
and $\phi_1^2+\phi_2^2+\phi_3^2 = 1$. \cite{chern06}
Note that the phase factor $e^{i\,\mathbf Q_0\cdot\mathbf r} = \pm 1$ changes sign between consecutive layers
of tetrahedra along the $z$ direction (cf. the two layers of green tetrahedra in Fig.~\ref{fig:spiral}).

The calculation can be further simplified in the $J \to \infty$ limit.  The vanishing of the total magnetization on tetrahedra
in this limit requires $\hat\mathbf n_3 \parallel \partial_y\hat\mathbf n_1$, and the $\phi$ fields can be analytically expressed
in terms of the director fields $\hat\mathbf n_i$. The effective energy functional in the gradient approximation
can be solely expressed in terms of the director fields $\hat\mathbf n_1$
and $\hat\mathbf n_2\parallel \partial_y\hat\mathbf n_1\times\hat\mathbf n_1$~\cite{chern06}.
It contains two contributions:  $\mathcal{F} = \int d\mathbf r \left( E_K + E_{DM} \right)$.
The first term, originating from magnetoelastic coupling, gives penalties to variations
of the staggered magnetizations:
\begin{equation}
	\label{eq:EK}
	E_{K} = \frac{K_u}{4} \Bigl[\left(\partial_x\hat\mathbf n_1\right)^2 + \left(\partial_y\hat\mathbf n_1\right)^2
	+ 2\left(\partial_z\hat\mathbf n_1\right)^2 - \left(\hat\mathbf n_2\cdot\partial_z\hat\mathbf n_1\right)^2\Bigr], \nonumber \\
\end{equation}
while the DM term contains Lifshitz invariants which are first-order in the gradients of $\hat\mathbf n_1$:
\begin{equation}
	\label{eq:ED}
	E_{DM} = -D\hat\mathbf n_1\cdot\bigl(\hat\mathbf a\times{\partial_x\hat\mathbf n_1}
	+\hat\mathbf b \times {\partial_y \hat\mathbf n_1}
	-\hat\mathbf c \times {\partial_z \hat\mathbf n_1}\bigr)/4. \quad
\end{equation}
Here we orient  the axes of coordinate reference frame $x,y, $ and $z$  along the principal axes $a,b$, and $c$ of the crystal.

The form of the DM functional suggests spiral solutions with $\hat\mathbf n_1$ rotating about one of the principal
axes and staying in the plane perpendicular to it. These special solutions were first obtained in Ref.~\onlinecite{chern06}.
For example, the solution $\hat\mathbf n_1 = (\cos\theta(y), 0, \sin\theta(y))$ with $\theta(y) = 2\pi\delta y + \psi$
describes a helical order with coplanar spins lying in the $ac$ plane (see Fig.~\ref{fig:spiral}).
This spiral magnetic order producing a Bragg scattering at $\mathbf Q = 2\pi(0, \delta, 1)$
is consistent with the experimental characterization of CdCr$_2$O$_4$. \cite{chung05}

The energy functional $\mathcal{F}$ actually possesses a continuous symmetry related to the O(2) rotational
invariance of the helical axis. To describe the general coplanar spiral solution,
we first introduce two orthogonal unit vectors lying in the $xy$ plane:
\begin{eqnarray}
\hat{\bm \xi} &=& +\cos \Phi \,\hat{\mathbf x} + \sin \Phi \,\hat{\mathbf y}, \nonumber \\
\hat{\bm\zeta} &=& -\sin \Phi \,\hat{\mathbf x}+ \cos \Phi \,\hat{\mathbf y}
\end{eqnarray}
A coplanar spiral with spins rotating about the $\hat{\bm\xi}$ direction
has the form
\begin{equation}
\label{eq:general_n1}
\hat\mathbf n_1(\mathbf r) = \cos\theta(\xi)\,\hat{\bm\zeta} + \sin\theta(\xi)\,\hat\mathbf z, \qquad \xi \equiv \mathbf r\cdot\hat{\bm\xi}.
\end{equation}
Substituting $\hat\mathbf n_1(\mathbf r)$ into the Eqs.~(\ref{eq:EK})
and (\ref{eq:ED}) yields an energy density:
$
	E[\theta] = -D \partial_{\xi}\theta +\frac{1}{4} K_u \left(\partial_\xi\theta\right)^2,
$
whose minimization gives a helical wavenumber
\begin{eqnarray}
	\mathcal Q = 2\pi\delta = d\theta/d\xi = 2D/K_u.
	\label{eq:delta}
\end{eqnarray}
The O(2) invariance indicates that the magnetic energy
is independent of the angle $\Phi$, i.e. $\delta \mathcal F[\hat{\mathbf n}_1] / \delta \Phi = 0$.
It is worth noting that this degeneracy is similar to the O(3) symmetry for the helical axis
observed in the spiral order of MnSi.~\cite{belitz06}
More importantly, as will be discussed below, it is exactly this additional symmetry that gives rise to
the anisotropic dispersion of helimagnons.

The coplanar spirals described above (with its axis lying in the $xy$ plane) are also
accidentally degenerate with a spiral in which spins rotate about the $z$ axis.~\cite{chern06}
In real compounds, this accidental degeneracy as well as the aforementioned O(2) symmetry are
lifted by further neighbor interactions and cubic symmetry field of the lattice.
In particular, the further-neighbor exchange contributes to a gradient term $(J_3 - J_2) (\partial_z \hat{\mathbf n}_1)^2$
in the gradient expansion.~\cite{chern06}
The experimentally observed $(0,\delta,1)$ spiral is selected by a large antiferromagnetic third-neighbor
exchange $J_3$ which favors coplanar spirals rotating about the $x$ or $y$ axes.
The explicit magnetizations of a spiral with the pitch vector parallel to the $y$-axis are
\begin{eqnarray}
	{\bar \mathbf S}_\mu(\mathbf r) = \hat\mathbf a\,\sin(\mathbf Q\cdot\mathbf r+\varphi_{\mu})
	+ \hat\mathbf c\,\cos(\mathbf Q\cdot\mathbf r+\varphi_{\mu}),
	\label{eq:S_bar}
\end{eqnarray}
where $\mu = 0,1,2,3$ is the sublattice index, $\varphi_0=\psi$, $\varphi_1=\pi\delta + \psi$, $\varphi_2=\pi + \psi$,
and $\varphi_3=\pi(1+\delta) + \psi$, with $\psi$ an arbitrary constant related to the U(1) symmetry of the coplanar spiral,
i.e. $\theta(\mathbf r) = \mathbf Q\cdot\mathbf r + \psi$.

It is worth noting that the pitch of magnetic spirals induced by DM interaction is usually rather long with
the wavenumber of the order $\delta \sim D/J \sim 10^{-3}-10^{-2}$.
The relatively short pitch of spirals in CdCr$_2$O$_4$ is due to
the ineffectualness of $J$ in stabilizing the magneitc order. Instead, the DM interaction competes
with the magnetoelastic coupling and the helical wavenumber is given by $\delta \sim D/K \approx 0.1$.

\subsection{Spherical approximation}
\label{sec:spherical}

While the above continuum theory gives a rather consistent and simple picture of helical magnetic orders
in tetragonal pyrochlore antiferromagnets, it is based on the assumption of an infinite $J$.
To go beyond this limit and investigate the magnetic orders when the various parameters are of similar order,
we resort to the spherical approximation, in which the local length constraints $|\mathbf S_i| = S$ are replaced
by a global one $\sum_{i=1}^N |\mathbf S_i|^2 = NS^2$, where $N$ is the number of lattice sites.
With this soft constraint, the model Hamiltonian~(\ref{eq:H_model}) can be diagonalized with the aid
of Fourier transform $\mathbf S_i = \sum_{\mathbf k} \mathbf S_\mu(\mathbf k)\,e^{i\mathbf k\cdot\mathbf r}$.
Here we label a lattice site as $i = (\mu, \mathbf r)$, where $\mu = 0, 1, 2, 3$ is the sublattice index and $\mathbf r$
denotes its position. The Hamiltonian then becomes
\begin{eqnarray}
	\mathcal{H} &=& \sum_{\mathbf k} \sum_{\mu,\nu} J_{\mu\nu}(\mathbf k)\,
	\mathbf S_\mu(\mathbf k)\cdot\mathbf S_\nu(-\mathbf k)
	\nonumber \\
	&+& \sum_{\mathbf k} \sum_{\mu,\nu}\sum_{\alpha,\beta}\, D^{\alpha\beta}_{\mu\nu}(\mathbf k)
	S^\alpha_\mu(\mathbf k)\,S^\beta_\nu(-\mathbf k),
	\label{eq:H-spherical}
\end{eqnarray}
where the matrices $J_{\mu\nu}(\mathbf k)$ and $D^{\alpha\beta}_{\mu\nu}(\mathbf k)$ are the Fourier transform
of the exchange and DM interactions, respectively. The explicit form of the matrices are given in the Appendix \ref{a0}.
The ground-state energy is given by the minimum eigenvalue $\lambda_{\rm min}$ of the
matrix $H^{\alpha\beta}_{\mu\nu} = J_{\mu\nu}\delta^{\alpha\beta}
+ D^{\alpha\beta}_{\mu\nu}$. In particular, the ordering wavevector $\mathbf Q$ is obtained by minimizing $\lambda_{\rm min}(\mathbf k)$
with respect to $\mathbf k$, i.e. $E_0 = N \lambda_{\rm min}(\mathbf Q)$.

\begin{figure}
\begin{centering}
\includegraphics[width=1\columnwidth]{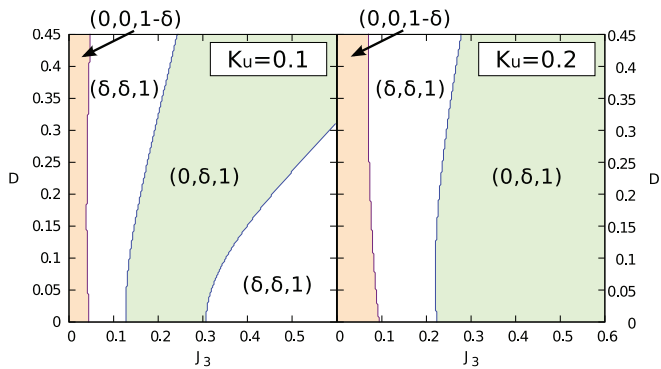} 
\par\end{centering}
\caption{(Color online) Phase diagram of helical magnetic orders on pyrochlore lattice
based on spherical approximation. Both $D$ and $J_3$ are measured with respect to the NN exchange $J$,
and we have set the odd-parity exchange anisotropy $K_u = 0.1J$ and $0.2J$ while leaving the even-parity exchange anisotropy $K_g = 0$. 
The green, yellow, and white regions are characterized by
a ordering wavevector $\mathbf Q = 2\pi(0, \delta, 1)$, $2\pi(0, 0, 1-\delta)$, and $2\pi(\delta, \delta, 1)$,
respectively.
\label{fig:phase}}
\end{figure}

Fig.~\ref{fig:phase} shows the ordering wavevector $\mathbf Q$
as a function of DM interaction and third-neighbor coupling $J_3$. We find three different helical orders characterized
by an incommensurate wavenumber $\delta \sim D/K_u$.
At small $J_3$, the ordering wavevector $\mathbf Q = 2\pi (0, 0, 1-\delta)$ indicates a magnetic spiral in which
spins rotate about the $z$ axis. This spiral solution has been discussed in Ref.~\onlinecite{chern06} and is accidentally
degenerate with the coplanar spirals described in Eq.~(\ref{eq:S_bar}) in the $J \to \infty$ limit.
A new spiral order with ordering wavevector $\mathbf Q = 2\pi (\delta, \delta, 1)$ is obtained for a small window of $J_3$ (depending
on the value of $K_u$). This spiral state with its axis pointing along the diagonal in the $xy$ plane is also
included in the general coplanar spiral solution Eq.~(\ref{eq:general_n1}) in the infinite $J$ limit.
As $J_3$ is further increased, the $(0, \delta, 1)$ spiral observed in CdCr$_2$O$_4$ becomes the ground state.
For a helical wavenumber $\delta \sim 0.1$, this spiral order is indeed the magnetic ground state for a $J_3 \approx 0.3J$,
consistent with the {\it ab initio} calculations. \cite{chern06,yaresko08}

Interestingly, for small exchange anisotropies, e.g. $K_u = 0.1 J$, the spiral order  $(0, \delta, 1)$
with axis along either $a$ or $b$ axes is sandwiched by the spiral state with axis along the $[110]$ direction,
and the ordering wavevector $\mathbf Q$ switches back to $2\pi(\delta, \delta, 1)$ for largest $J_3$ [Fig.~\ref{fig:phase}(a)].
On the other hand, for larger anisotropy $K_u$, the ground state remains the $(0, \delta, 1)$ spiral for large $J_3$.

\subsection{Perturbation correction to coplanar spirals}
\label{sec:perturbation}

We now focus on the helical magnetic order with wavevector $\mathbf Q = 2\pi(0, \delta, 1)$, which is relevant
for the case of CdCr$_2$O$_4$, and use perturbation method to obtain corrections $\mathbf s_i \sim \mathcal{O}(D/J)$
to the coplanar spiral solution Eq.~(\ref{eq:S_bar}) when the NN exchange is finite.
In the ground state, the magnetic moment is parallel or antiparallel to the effective exchange field
\begin{eqnarray}
	\label{eq:H-eff}
	\mathbf H_i = \partial\mathcal{H}/\partial\mathbf S_i = \sum_{j}  
	\left(J_{ij}\,\mathbf S_j - \mathbf D_{ij}\times\mathbf S_j\right),
\end{eqnarray}
such that the torque $\mathbf T_i = \mathbf S_i\times\mathbf H_i$ vanishes.
Mathematically, our perturbative calculation is based on a $1/J$ expansion of the equilibrium equation $\mathbf T_i = 0$.
The zeroth-order result corresponds to the $J \to \infty$ limit and is given by Eq.~(\ref{eq:S_bar}).
We express the exchange field as $\mathbf H_i = \bar{\mathbf H}_i + \mathbf h_i$,
the condition of zero torque requires
\[
	({\bar \mathbf S}_i + \mathbf s_i)\times({\bar \mathbf H}_i + \mathbf h_i) = 0.
\]
Collecting only terms linear in $D/J$ from this equation, we obtain a set of coupled equations:
\begin{eqnarray}
\label{eq1}
	& & \qquad\qquad\quad \sum_{j}J_{ij}\cos\theta_{ij}\,(s^{x}_i-s_{j}^{x})=0,  \\
\label{eq2}
	& & \!\!\!\!\! \sum_{j}\left\{D_{ij}^{x}\cos\theta_{j}-D_{ij}^{z}\sin\theta_{j}
	 - J_{ij}\left[\cos\theta_{ij}\,s_{i}^{y}-s_{j}^{y}\right]\right\} = 0,  \nonumber \\
\end{eqnarray}
where $\theta_{i}={\bf Q}\cdot{\bf r}+\psi_\mu$, and $\theta_{ij} = \theta_i - \theta_j$.
Sine the NN exchange $J$ can be expressed as $(J/2)|\mathbf M(\mathbf r)|^2$, the $J \to \infty$ limit imposes
hard constraints $\mathbf M(\mathbf r) = 0$ on all tetrahedra. A finite value of $J$ is thus expected
to introduce a finite total magnetization.
For simplicity, here we consider anisotropies dominated by a tetragonal distortion with odd parity only.
Solving the above equations yields ($\mu = 0, \cdots, 3$)
\begin{eqnarray}
	\label{eq:1st-order}
	\mathbf s_\mu(\mathbf r) =  \frac{-\sqrt{2}D}{(4J+10K_u)}\,\cos\left(\mathbf Q\cdot\mathbf r\right)\,\hat\mathbf b.
\end{eqnarray}  
Interestingly, the first-order corrections does not modify the spiral pitch; the incommensurate ordering wavevector
is still given by the zeroth order expression 
$\mathcal{Q}=2\pi\delta$.
As the correction is uniform for the four sublattices, the finite value of $J$ gives rise to a modulated ferromagnetic
component $\mathbf M(\mathbf r)$ on individual tetrahedron.

\begin{figure}
\centering{}
\includegraphics[width=0.98\columnwidth]{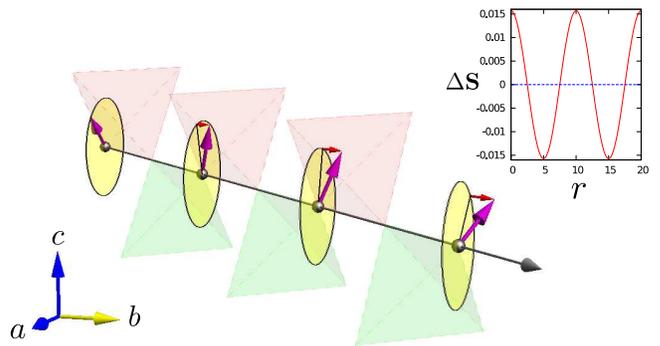}
\caption{(Color online) A schematic diagram showing the out-of-plane corrections to a coplanar helical order.
The coplanar spins are described by a vector field $\hat\mathbf n_1 = (\cos\theta(y), 0, \sin\theta(y))$,
with $\theta(y) = 2\pi\delta y + \mbox{const}$, while the out-of-plane components are given by Eq.~(\ref{eq:1st-order}).
The inset shows the out-of-plane spin component $\Delta\mathbf S$ as a function of $r$ computed for the following set of
parameters: $J=1.35\mbox{~meV}$, $K_{u}=0.21\mbox{~meV}$, $K_{g}=0\mbox{~meV}$, $D=0.14\mbox{~meV}$,
and $J_{3}=0.28\mbox{~meV}$.\label{fig:correction} } 
\end{figure}

\subsection{Relaxation dynamics simulations}
\label{sec:relaxation}

While the perturbation correction gives a better approximation of the helical order in pyrochlore lattice,
a high-precision description of the magnetic structure is required for the calculation of spin-wave excitations.
A common approach to obtain the magnetic ground state is the low-temperature Monte Carlo simulations.
However, the spin configurations obtained from Monte Carlo simulations are still too noisy for the purpose of
 the  calculation of magnon spectrum.
Instead, here we resort to the overdamped dynamics simulations based on the Landau-Lifshitz-Gilbert (LLG) equations.
In particular, since the coplanar spiral solution with the perturbation corrections already provide a close
approximation to the true ground state, such relaxation simulations can efficiently bring the system to
the desired magnetic order.

\begin{figure}
\begin{centering}
\includegraphics[width=0.9\columnwidth]{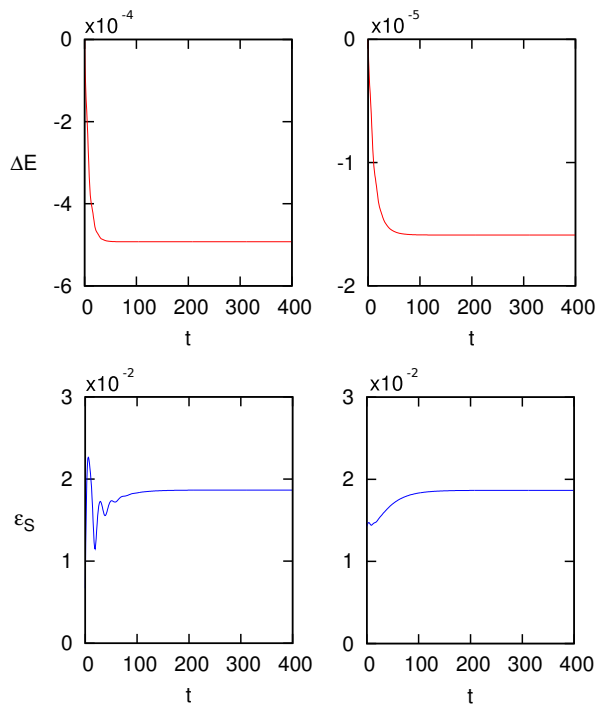} 
\par\end{centering}
\caption{(Color online) The energy decrease $\Delta E(t) = E(t) - E(t=0)$ as a function of simulation time
starting with 
(a) coplanar spiral solution Eq.~(\ref{eq:S_bar}) and (b)  initial state including 
first order correction~(\ref{eq:1st-order}). The energy and time are measured with respect to $J$ and $1/J$, respectively.
We performed the simulations on a lattice with $16\times 10^3$ spins with periodic boundary conditions.
Panels (c) and (d) show the corresponding averaged spin deviations $\epsilon_S(t)=\sum_{i}|{\bf S}_{i}(t)-{\bar\mathbf S}_{i}|/N$
from the coplanar state and the inclusion of the first order correction respectively as a function of simulation time. 
\label{fig:trace}}
\end{figure}

The LLG equation is a first-order differential equation describing the dynamics of classical magnetic moments:
\begin{eqnarray} 
	\label{eq:LLG}
	\frac{\partial\mathbf{S}_{i}}{\partial t}= \mathbf{S}_{i}\times \mathbf H_i +
	\frac{\alpha_{G}}{S}  \mathbf{S}_{i}\times \frac{\partial \mathbf{S}_{i}}{\partial t},
\end{eqnarray}
where $\mathbf H_i$ is the effective exchange field defined in Eq.~(\ref{eq:H-eff}), and $\alpha_{G}$ is a
dimensionless damping parameter. We employ a finite-difference method to integrate the LLG equation
numerically. \cite{serpico01} The technical details of the numerical method can be found in Appendix \ref{a1}.
We performed our simulations on a cluster of $10^3$ cubic unit cells with $N = 16000$ spins.
Periodic boundary conditions were used in all simulations. We have also tested our algorithms on some well studied
Hamiltonians and obtained excellent agreement with exact solutions.

\begin{figure}
\begin{centering}
\includegraphics[width=0.92\columnwidth]{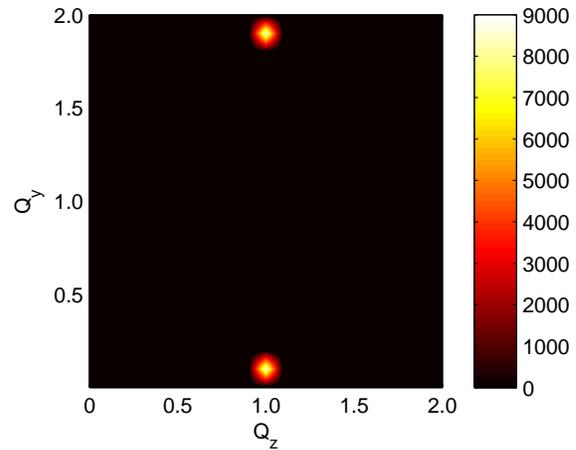}
\par\end{centering}
\caption{(Color online) Structure factor of the helical spin order obtained from the relaxation dynamics
simulation. Sharp peaks appear at the ordering wavevector ${\bf Q} = 2\pi (0, \delta, 1)$.
\label{fig:s-factor}}
\end{figure}

We start our dynamics simulation from the coplanar spiral state both with and without the first-order corrections.
Fig.~\ref{fig:trace} shows the simulation results for parameters $K_{u}=0.1J$, $D=-0.069J$ and $J_3=0.2J$; 
the corresponding helical wavenumber is $\delta = 0.1$.
In both cases, the energy monotonically decreases with simulation time, indicating a steady relaxation towards
the true helical magnetic ground state.  Also shown are the averaged deviations of
spin configurations $\mathbf S_i(t)$ from the coplanar helical state (with and without first-order correction)
as a function of simulation time.
The rather small deviations of the order $10^{-2}$ indicate that both the coplanar spiral
and the first-order result are indeed already good approximations to the ground state.
Fig.~\ref{fig:s-factor} shows the structure factor
$\mathcal{S}(\mathbf k) = \left|(1/N)\sum_i\mathbf S_i \exp({i\mathbf k\cdot\mathbf r_i})\right|^2$
of the spin configurations obtained from the relaxation dynamics simulations.
As expected, the structure factor exhibits peaks at the ordering wavevector $\mathbf Q = 2\pi(0, \delta, 1)$.

\section{Spin-wave excitations}

In this Section we study magnetic excitations of the helical magnetic order on the pyrochlore lattice.
We first discuss the so-called helimagnons in the continuum limit. The appearance of this special gapless mode
originates from the unique symmetry-breaking in chiral magnets.
However, since the spiral pitch in CdCr$_2$O$_4$ is relatively short compared with helical orders
in other systems, the crystal effects are expected to modify the helimagnon dispersion significantly.
We thus also compute the spin-wave dispersions based on microscopic spin models. Starting from the
high-precision ground state obtained in the previous section, we performed exact diagonalizations on a
large unit cell to study the nature of the low-energy acoustic-like modes. The results are then compared with
the predictions based on helimagnon theory. We also employed dynamics simulations on finite systems
to obtain the full magnon spectrum and compare the results with the experiments.

\subsection{Helimagnons in the continuum limit}
\label{sec:goldstone}

We first consider magnetic fluctuations in the helical order within the framework of the continuum theory.
For a spiral with a fixed axis, the obvious soft modes are phase fluctuations associated with the
coplanar director field. More specifically, let us consider a spiral with its axis along the $y$-direction:
$\hat\mathbf n_1 = (\cos\theta, 0, \sin\theta)$. Introducing a slowly-varying perturbation to the
phase angle $\theta = \mathcal Q y + \psi(\mathbf r)$, the energy functional associated with the fluctuations
is $\mathcal{E}[\psi] = K_u \int d\mathbf r \left[\nabla\psi(\mathbf r)\right]^2$. This indeed describes a soft mode
similar to that of a planar magnet.  However, as is well known in chiral nematics and smectic liquid crystals,
a phase fluctuation of the form $\psi = \alpha \,\zeta$ corresponds to a simple rotation of
the helical axis. \cite{chaikin}
Since coplanar spirals with different axes are energetically degenerate (see discussions in Sec.~\ref{sec:continuum}),
such rotations cannot cost energy, yet $\mathcal{E} \sim \alpha^2 \neq 0$ for this particular
phase fluctuation. Consequently, there cannot be any $(\partial_\zeta\psi)^2$ term in the effective energy density.
The correct energy functional thus has the form \cite{chaikin,belitz06}
\begin{eqnarray}
	\mathcal{E} = \frac{1}{2}\!\int\! d\mathbf r \Bigl[c_z\!\left(\partial_z \psi\right)^2
	+\!c_{\parallel}\! \left(\partial_{\xi}\psi\right)^2
	+\! \frac{c_{\perp}}{\mathcal Q^2}\! \left(\partial^2_{\zeta} \psi\right)^2 \!\Bigr], \,\,
\end{eqnarray}
where $c_\parallel$, $c_{\perp} \sim K_u$, $c_z \sim J_3$ are elastic constants, and $\mathcal Q = 2\pi\delta$ is the pitch wavenumber.
Note that the first term in the above equation arises from a rather large $J_3$ which gives a penalty to variations along the $z$ direction.

A derivation of the dynamics equations starting from the microscopic model~(\ref{eq:H_model})
is tedious and almost intractable. To obtain the dispersion of the soft mode, it is more illuminating to employ
a simple phenomenological approach based on the time-dependent Ginzburg-Landau theory. \cite{chaikin,belitz06}
In this formalism, the Goldstone mode $\psi$ couples to a mode $m$ at zero wavevector, which is soft due to spin conservation.
Schematically the coarse-grained magnetization field is
\begin{eqnarray}
	\label{eq:Sr}
	& &\mathbf S_\mu(\mathbf r) \sim {\bar \mathbf S}_\mu(\mathbf r) \, + \, m(\mathbf r) \hat\mathbf y \\
	& & \quad + \,\,\psi(\mathbf r)\,\bigl[\cos (\mathbf Q\cdot\mathbf r + \varphi_\mu)\,
	\hat\mathbf x - \sin (\mathbf Q\cdot\mathbf r + \varphi_\mu)\, \hat\mathbf z\bigr], \nonumber
\end{eqnarray}
where ${\bar \mathbf S}_\mu$ is the coplanar
spiral given in Eq.~(\ref{eq:S_bar}) and $\varphi_\mu$ denotes the relative phase shift of the sublattices.
The total energy functional becomes
\begin{eqnarray}
	\mathcal{F}[m, \psi] = \frac{r_0}{2}\int d\mathbf r \, m^2(\mathbf r) + \mathcal{E}[\psi].
\end{eqnarray}
Here $r_0$ is an effective mass of the zero-wavevector mode.
The magnetization dynamics is governed by a generalized Landau-Lifshitz equation:
\begin{eqnarray}
	\partial_t \mathbf S_\mu(\mathbf r) = -\gamma\, \mathbf S_\mu(\mathbf r) \times
	\frac{\delta \mathcal{F}}{\delta \mathbf S_\mu(\mathbf r)}
\end{eqnarray}
where $\gamma$ is a constant. Substituting (\ref{eq:Sr}) into the above equation and following similar
steps outlined in Ref.~\onlinecite{belitz06}, we arrive at the following coupled equations
\begin{eqnarray}
	\partial_t m(\mathbf r) = -\gamma \frac{\delta \mathcal{F}}{\delta \psi(\mathbf r)}, \qquad
	\partial_t \psi(\mathbf r) = \gamma \frac{\delta \mathcal{F}}{\delta m(\mathbf r)}.
\end{eqnarray}
To obtain the spectrum of the soft mode, we eliminate $m$ in the above equations to obtain a wave equation
for the $\psi$ field
\begin{eqnarray}
	\partial_t^2\psi(\mathbf r)
	= -r_0 \gamma^2 \left(-c_z \partial_z^2 -c_{\parallel} \partial_\xi^2
	+ \frac{c_{\perp}}{\mathcal Q^2} \partial_{\zeta}^4\right) \psi(\mathbf r).
\end{eqnarray}
Assuming a space-time dependence $\psi \sim \exp(i\mathbf k\cdot\mathbf r - i\epsilon_{\mathbf k} t)$, the energy of the
soft mode is
\begin{eqnarray}
	\label{eq:omega}
	\epsilon(\mathbf k) = \gamma \, r_0^{1/2}\, \sqrt{c_z k_z^2 + c_{\parallel} k_{\parallel}^2
	+ \frac{c_{\perp}}{\mathcal Q^2} k_{\perp}^4}.
\end{eqnarray}
The dispersion is highly anisotropic: for wavevectors parallel to the spiral $\xi$-axis direction or the $z$-axis the dispersion is linear
as in an antiferromagnet, while it is quadratic for wavevectors parallel to the $x$-axis.

\begin{figure*}
\centering{}
\includegraphics[width=1.7\columnwidth]{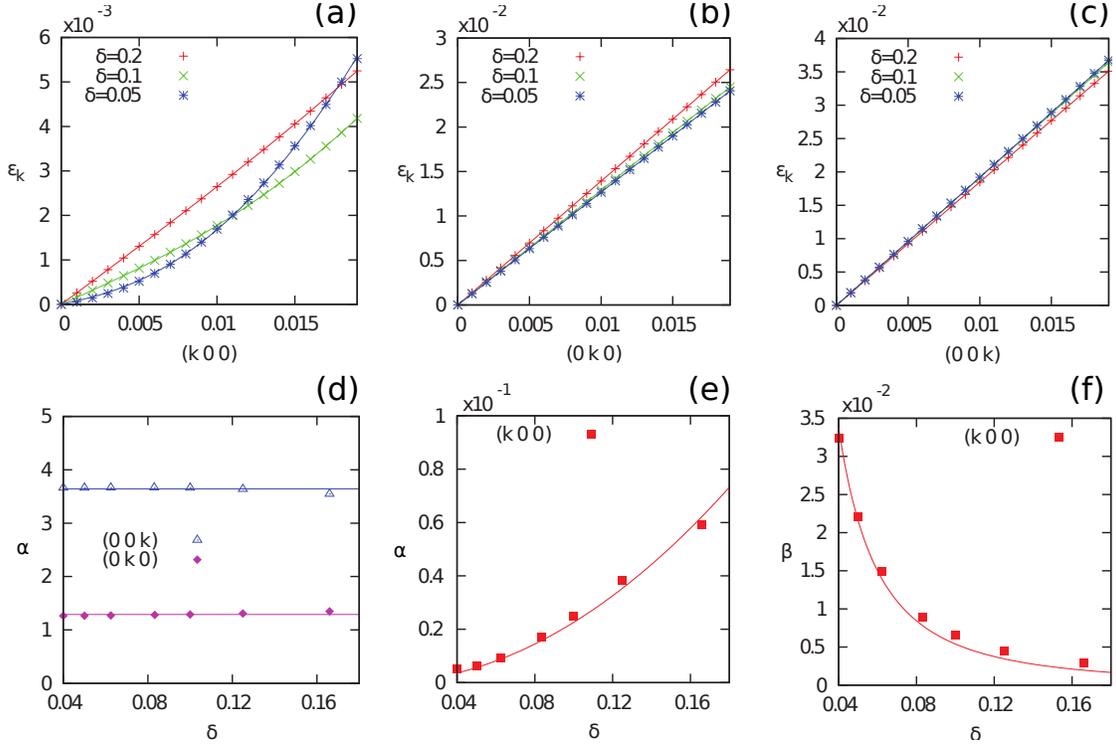}
\caption{(Color online) Spin-wave dispersions along three principal axes [panels (a)--(c)] obtained from numerical
exact diagonalization of a large unit cell. The results are fitted to the phenomenological
dispersion $\epsilon_{\mathbf k} = \sqrt{\alpha k^2 + \beta k^4}$. The magnon energy is measured in unit of $JS$.
Panels~(d)--(f) show the dispersion parameters $\alpha$ and $\beta$ as a function of the
helical wavenumber $\delta$. These data are fitted to $\alpha=$const, $\alpha \propto \delta^2$,
and $\beta\propto \delta^{-2}$, respectively.
\label{fig:helimagnon}}
\end{figure*}

\subsection{Exact diagonalization with large unit cell}

As discussed in Sec.~\ref{sec:continuum}, the helical wavenumber in tetragonally distorted pyrochlore is
determined by the ratio of DM interaction to exchange anisotropy (instead of the NN exchange constant).
The spiral pitch in CdCr$_2$O$_4$ is relatively short compared with that in other helical orders.
We thus expect significant anisotropy coming from the microscopic crystal fields. In addition, the $(0,\delta,1)$
spiral which is relevant to our case is stable only in the presence of a small third-neighbor coupling.
A gradient-expansion analysis shows that spin stiffness along the $z$ axis is increased in the presence of $J_3$.
To address these issues, here we perform exact diagonalizations with a large unit cell to investigate the
low-energy acoustic-like magnons and compare the results with predictions based on the helimagnon theory.

To begin with, we first choose a commensurate unit cell composed of $\Lambda$ cubic unit cells along the b axis (each cube contains 4 tetrahedra),
corresponding to an ordering wavevector $\mathbf Q = 2\pi(0, \delta, 1)$ with $\delta = 1/\Lambda$.
The DM constant that is required to stabilize such a spiral is then determined
using the expression
\begin{equation}
D = -(3/\sqrt{2})K_u\tan\pi\delta  
\end{equation}
This is the discrete version of Eq.~(\ref{eq:delta}) for a given fixed anisotropy $K_u$, and is obtained
by substituting the coplanar spiral ansatz (\ref{eq:S_bar}) into Hamiltonian (1) and minimizing the resulting expression
with respect to $\delta$. The relaxation dynamics simulation discussed in Sec.~\ref{sec:relaxation} is carried out to obtain an 
accurate description of the ground state.
We then introduce small deviations to spins in the ground state: $\mathbf S_i = \mathbf S_i^0 + \bm\sigma_i$,
where $\bm\sigma_i \perp \mathbf S_i^0$ denotes transverse deviations. Note that because of this constraint,
there are two degrees of freedom associated with $\bm\sigma_i$ at each site.
The LLG equation is linearized with respect to the small deviations $\bm\sigma_i$:
\begin{eqnarray}
	\label{eq:linear_LLG}
	\frac{\partial \bm\sigma_i}{\partial t}
	= \mathbf S_i^0\times\sum_j\left(J_{ij}\bm\sigma_j - \mathbf D_{ij}\times \bm\sigma_j\right)
	- \mathbf H_i^0 \times \bm\sigma_i,
\end{eqnarray}
where $J_{ij}$ includes $J$, $K_{ij}$ and further-neighbor exchanges, $\mathbf H_i^0 =\sum_{j} (J_{ij}\mathbf S_j^0
- \mathbf D_{ij}\times \mathbf S_j^0 )$ is the equilibrium local exchange field,
and we have set the damping constant $\alpha_G$ to zero.
The eigenmodes of the LLG equation has the form $\bm\sigma_i = \bm\sigma_m\,\exp({i\mathbf k\cdot\mathbf r-i\epsilon_{\mathbf k} t})$,
where the site index $i = (\mathbf r, m)$, where $\mathbf r$ denotes the position of the extended unit cell
containing $\Lambda$ cubes and $m$ labels spins within this block. Note that the above eigenmode $\{\bm\sigma_m\}$
has a momentum $\mathbf Q + \mathbf k$ due to the underlying structure of the supercell.
Arranging $\bm\sigma_m = (\sigma_m^x, \sigma_m^y)$
into a column vector $\vec {\mathbf u} = (\sigma_1^x, \sigma_1^y, \cdots, \sigma_s^x, \sigma_s^y)$
of dimension $n = 2\times 4\Lambda \times 4$, the linearized LLG equation is recast into an eigenvalue
problem $\hat{\mathbf T}(\mathbf k)\cdot\vec{\mathbf u} = -i\epsilon_{\mathbf k}\, \vec{\mathbf u}$.
Numerical diagonailization of the $n\times n$ matrix $\hat{\mathbf T}(\mathbf k)$ gives
the dispersion of the low-energy magnons in the vicinity of ordering wavevector $\mathbf Q$.

The resulting magnon dispersions along the three principal directions are shown in Fig.~\ref{fig:helimagnon}(a)--(c).
The number of spins in the unit cell is defined by the value of the spiral pitch $\delta$.
We have considered $\delta =0.2$, 0.1, 0.05, which gives $N_s = 80$, 160, 320, respectively.
The numerical spectra are fitted to a phenomenological dispersion: $\varepsilon_{\mathbf k} = \sqrt{\alpha k^2 + \beta k^4}$.
The dispersion parameters $\alpha$ and $\beta$ obtained from the fitting are shown in Fig.~\ref{fig:helimagnon}(d)--(f)
as a function of the helical wavenumber $\delta$.
First, we note that the dispersion along the $y$ (helical direction) and $z$ (staggering direction)
axes can be well approximated by a linear relation, $\varepsilon_{\mathbf k} \sim v k$, with $v = \sqrt{\alpha}$
weakly dependent on $\delta$, see Fig.~\ref{fig:helimagnon}(d). The large spin-wave velocity along the $z$ axis can be attributed
to a large $J_3$, consistent with the analysis of gradient expansion.~\cite{chern06}

On the other hand, the spectrum along the $x$-axis exhibits a predominant
quadratic behavior at small $k$, similar to that of a ferromagnet. Although the continuum theory predicts
an exact quadratic dispersion along the $x$ direction, Eq.(\ref{eq:omega}), careful fitting shows a small linear component $\alpha$,
which can be attributed to the discrete cubic symmetry of the underlying lattice model.
Dimensional analysis indicates that $\alpha$ should scale as $\mathcal Q^2$,~\cite{belitz06} while the
coefficient $\beta \sim 1/\mathcal Q^2$ according to the helimagnon dispersion Eq.~(\ref{eq:omega}).
These scaling relations are indeed confirmed by our exact diagonalization as demonstrated
in Figs.~\ref{fig:helimagnon}(e) and  (f).

\section{Magnon spectrum of $\boldsymbol{\mathrm{CdCr_{2}O_{4}}}$}
\label{sec:magnon}

\subsection{Linear dynamics simulation}
\label{sec:dynamics}

Although the exact diagonalization method discussed above provides a rather accurate magnon spectrum,
the complicated band-folding due to large unit cell makes it difficult to compare the results with experiment.
Instead, here we employ a less accurate but more direct approach based on simulating the real-time dynamics
of spins in a large finite lattice. Specifically we introduce a local basis for the transverse
fluctuations: $\bm\sigma_i = \sigma^x_i\,\hat\mathbf e^x_i + \sigma^y_i\,\hat\mathbf e^y_i$,
where $\hat\mathbf e^\alpha_i$ are two unit vectors orthogonal to the equilibrium spin direction $\hat\mathbf n_i$.
Substituting $\bm\sigma_i(t)$ into the linearized LLG equation yields
\begin{eqnarray}
	\begin{array}{l}
	\partial_t\sigma_{i}^{x}
	=\left(\tau_{i}^{x}-\alpha_{G}\,\tau_{i}^{y}\right)/(1+\alpha_G^2) + \zeta^x_i(t),
	\\ \\
	\partial_t\sigma_{i}^{y}
	=\left(\alpha_{G}\,\tau_{i}^{x}+\tau_{i}^{y}\right)/(1+\alpha_G^2) + \zeta^y_i(t),
	\end{array}
	\label{eq:LLG-sigma}
\end{eqnarray}
where $\tau_{i}^{x}$ and $\tau_{i}^{y}$ are linearized torques projected onto the local basis;
see Appendix~\ref{a2} for details of the calculations. A small damping is added to ensure stable simulations.
The dynamics simulation is initiated by a short pulse $\zeta_i(t)=\zeta_0 \exp(-t^2/{w}^2)\,\delta_{i, 0}$
localized at $\mathbf r_i = 0$. Here $\zeta_0$ determines the strength of the initial perturbation and $w$ is the width of the pulse.

\begin{figure}
\begin{centering}
\includegraphics[width=8cm]{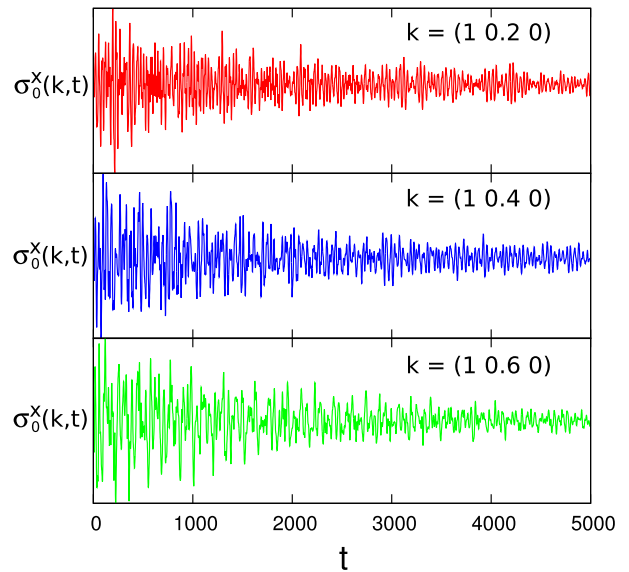}
\par\end{centering}
\caption{(Color online) Time evolution of $\sigma^x_0({\bf k},\, t)$ for different values of ${\bf k}$.
\label{fig:sigmakt}}
\end{figure}

The differential equations (\ref{eq:LLG-sigma}) are integrated with the aid of the fourth-order Runge-Kutta method.
The solution $\bm\sigma_i = \bm\sigma_{m}({\bf r},\, t)$ describes the evolution of each individual spin due
to the localized short pulse. Since the system is driven by a white source with flat spectrum, we expect magnetic excitations
of various energy and momentum are generated in our simulations. The spin-wave spectrum is then obtained
via the Fourier transform of the simulation data. First we performed the spatial Fourier transform
$\bm\sigma_{m}({\bf k},\, t)=\sum_{\mathbf r}\bm\sigma_{m}({\bf r},\, t)e^{i {\bf k}\cdot{\bf r}}$, where
only those momenta ${\bf k}$ which satisfy the periodic boundary conditions of the finite lattice are allowed.
In Fig. \ref{fig:sigmakt} we present  the time evolution of  $\sigma_{m}({\bf k},\, t)$ for different values of ${\bf k}$. 
The magnon spectrum is then obtained the temporal Fourier transformation:
$\bm\sigma_{m}({\bf k},\, \epsilon)=\int  e^{i \epsilon t}\bm\sigma_{m}({\bf k},\, t)\,dt$.
The spectrum obtained for a set of model parameters that are realistic for $\mathrm{CdCr_{2}O_{4}}$
is discussed in the next subsection.

\subsection{Comparison with experiments}

\begin{figure}
\begin{centering}
\includegraphics[width=8cm]{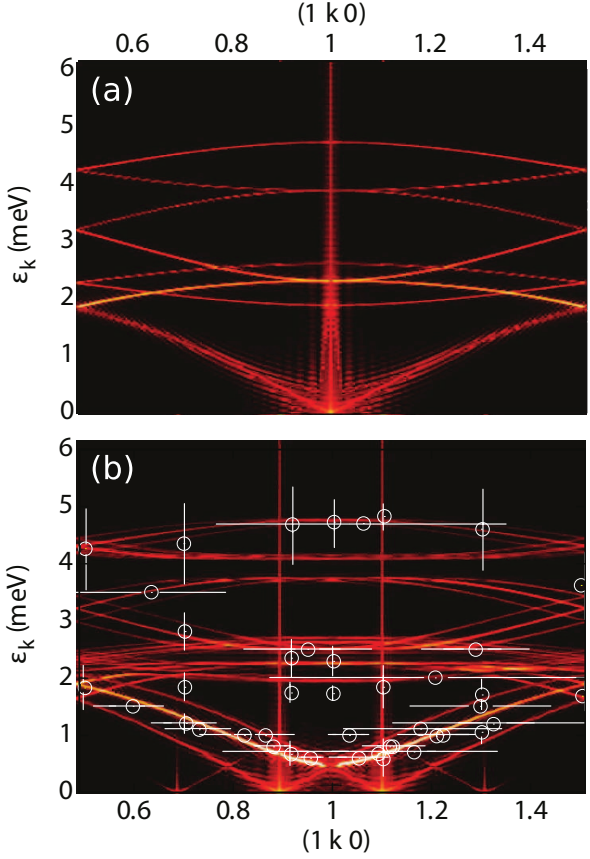}
\par\end{centering}
\caption{(Color online) (a) Spin wave spectrum of a N\'eel state with wavevector $\mathbf q = 2\pi(0,0,1)$.
This collinear order is the ground state of $\mathcal{H}$ with
parameters $J=1.35$ meV, $K_{u}=0.21$ meV, $D=0$ meV, and $J_{3}=0.28$ meV.
(b) Fitted spin wave spectrum to the neutron scattering measurement(white circles and rectangles)\cite{chung05}.
The final set of parameters are $J=1.35$ meV, $K_{u}=0.21$ meV, $D=0.14$ meV, and $J_{3}=0.28$ meV. 
Note that adding DM interaction lifts the gap at $\mathbf{q}=2\pi(1,1,0)$ and splits each band into three.
All the parameters are determined based on this fitting. Note that vertical peaks at the ordering
wave vector $\mathbf{q}=2\pi(1,1\pm\delta,0)$ are numerical artifacts.
\label{fig:fitting}}
\end{figure}

The magnon spectrum  of $\mathrm{CdCr_{2}O_{4}}$ has been studied by Chung \textit{et al.}
using inelastic neutron scattering.\cite{chung05} Here we present our numerical calculations
and compare the results with the experimental findings.
The fact that the helical wavenumber $\delta \approx 0.09$ can be used to fix the ratio of the DM interaction
to the exchange anisotropy $K_u$. On the other hand, values of $K_u$ and $K_g$ can be estimated by the magnon
energies at the zone center, and we find that even tetragonal distortion does not play a significant role here.
As for further-neighbor exchanges, {\it ab initio} calculations find a negligible $J_2$ and a quite large $J_3$
with antiferromagnetic sign. \cite{chern06,yaresko08} 
Using the following set of parameters: $J=1.46$ meV, $K_{u}=0.34$ meV, $K_{g}=0$ meV, $D=0.078$ meV, and $J_{3}=0.30$ meV,
we obtained very good agreement with the experimentally measured spectrum, as shown in Fig.~\ref{fig:fitting}.

\section{Conclusion and Discussion}
\label{sec:conclusion}

We have studied the spin-wave excitations in a special helical magnetic order on the pyrochlore lattice.
This spiral structure characterized by a wavevector $\mathbf Q = 2\pi(0, \delta, 1)$ is observed
in frustrated spinel CdCr$_2$O$_4$.
Motivated by the experiments, we assumed that magnetic frustration is completely relieved by a
tetragonal distortion which preserves the lattice translational symmetry.
We have employed various analytical approaches and numerical methods to investigate the interplay
of Dzyaloshinskii-Moriya interaction and further-neighbor couplings.
Our spin-wave calculation with parameters inferred from structural
data and {\em ab initio} calculations agrees well with the measured magnon spectrum.
Combined with a systematic analysis of the helical structure, our study helps clarify the underlying mechanisms
that stabilize the observed helical order and provides a quantitative description of the model Hamiltonian.

We have also shown that the continuum model of the spin spirals based on a gradient-expansion of the order parameter
exhibits a continuous symmetry similar to that of smectic liquid crystals. A Goldstone mode with highly anisotropic
dispersion results from the unique broken symmetry of the helical order. Although the emergent continuous symmetry
is broken by the crystal fields effects at the microscopic level, our exact diagonalization on a large
unit cell finds acoustic-like magnons which are reminiscent of the Goldstone mode discussed above.

Several recent experiments have established CdCr$_2$O$_4$ as the paradigmatic example of magnetic ordering
induced by spin-lattice coupling in geometrically frustrated magnets. Although an overall tetragonal distortion
with $a = b < c$ is observed below the ordering temperature, details of the lattice distortion remain to be
clarified experimentally. The issue at hand is whether the distortion breaks the lattice inversion symmetry.
As pointed out in a previous theoretical study,~\cite{chern06} the broken parity plays an important role in stabilizing
the observed helical order. This is because the lack of inversion symmetry endows the crystal structure
with chirality, which is transferred to the magnetic order through relativistic
spin-orbit interaction. Experimental evidence implying a broken parity in CdCr$_2$O$_4$ was recently reported in
infrared absorption spectra. \cite{kant09}  Our numerical calculation on spin-wave dispersions also indicates
a predominantly tetragonal distortion which breaks the inversion symmetry, giving further support to the
scenario that the helical magnetic order inherits its chirality from a lattice with broken parity.

\section*{Acknowledgements}
We thank J. Deisenhofer, C.~Price, O. Sushkov, O.~Tchernyshyov for stimulating discussions.
N.P. and E.C. acknowledge the support from NSF grant DMR-1005932.
G.W.C. is supported by ICAM and NSF Grant DMR-0844115.
N.P. and G.W.C. also thank the hospitality of the visitors program at MPIPKS,
where part of the work on this manuscript has been done.

\appendix

\begin{widetext}

\section{Spherical approximation} 
\label{a0}

The matrices $J_{\mu\nu}(\mathbf k)$ and $D^{\alpha\beta}_{\mu\nu}(\mathbf k)$ in the spherical approximation (Eq.~\ref{eq:H-spherical}) are defined as follows:

\begin{eqnarray}
{\hat J}(\mathbf k)=\left(\begin{array}{cccc}
2J_{3}\left(\tilde{c}{}_{x\bar{y}}+\tilde{c}{}_{x\bar{z}}+\tilde{c}{}_{yz}\right) & (2J-K_{u})c_{yz}+3K_{u}s_{yz} & (2J-K_{u})c_{x\bar{z}}+3K_{u}s_{x\bar{z}} & 2(J+K_{u})c_{x\bar{y}}\\ \\
(2J-K_{u})c_{yz}-3K_{u}s_{yz} & 2J_{3}\left(\tilde{c}{}_{xy}+\tilde{c}{}_{xz}+\tilde{c}{}_{yz}\right) & 2(J+K_{u})c_{xy} & (2J-K_{u})c_{xz}+3K_{u}s_{xz}\\ \\
(2J-K_{u})c_{x\bar{z}}-3K_{u}s_{x\bar{z}} & 2(J+K_{u})c_{xy} & 2J_{3}\left(\tilde{c}{}_{xy}+\tilde{c}{}_{x\bar{z}}+\tilde{c}{}_{y\bar{z}}\right) & (2J-K_{u})c_{y\bar{z}}+3K_{u}s_{y\bar{z}}\\ \\
2(J+K_{u})c_{x\bar{y}} & (2J-K_{u})c_{xz}-3K_{u}s_{xz} & (2J-K_{u})c_{y\bar{z}}-3K_{u}s_{y\bar{z}} & 2J_{3}\left(\tilde{c}{}_{x\bar{y}}+\tilde{c}{}_{y\bar{z}}+\tilde{c}{}_{xz}\right)
\end{array}\right), \quad
\end{eqnarray}

\begin{eqnarray}
{\hat D}(\mathbf k)=\sqrt{2}D\left(\begin{array}{cccccccccccc}
0 & 0 & 0 & 0 & -c_{yz} & -c_{yz} & 0 & c_{x\bar{z}} & 0 & 0 & 0 & c_{x\bar{y}}\\
0 & 0 & 0 & c_{yz} & 0 & 0 & -c_{x\bar{z}} & 0 & c_{x\bar{z}} & 0 & 0 & -c_{x\bar{y}}\\
0 & 0 & 0 & c_{yz} & 0 & 0 & 0 & -c_{x\bar{z}} & 0 & -c_{x\bar{y}} & c_{x\bar{y}} & 0\\
0 & c_{yz} & c_{yz} & 0 & 0 & 0 & 0 & 0 & -c_{xy} & 0 & -c_{xz} & 0\\
-c_{yz} & 0 & 0 & 0 & 0 & 0 & 0 & 0 & -c_{xy} & c_{xz} & 0 & c_{xz}\\
-c_{yz} & 0 & 0 & 0 & 0 & 0 & c_{xy} & c_{xy} & 0 & 0 & -c_{xz} & 0\\
0 & -c_{x\bar{z}} & 0 & 0 & 0 & c_{xy} & 0 & 0 & 0 & 0 & c_{y\bar{z}} & -c_{y\bar{z}}\\
c_{x\bar{z}} & 0 & -c_{x\bar{z}} & 0 & 0 & c_{xy} & 0 & 0 & 0 & -c_{y\bar{z}} & 0 & 0\\
0 & c_{x\bar{z}} & 0 & -c_{xy} & -c_{xy} & 0 & 0 & 0 & 0 & c_{y\bar{z}} & 0 & 0\\
0 & 0 & -c_{x\bar{y}} & 0 & c_{xz} & 0 & 0 & -c_{y\bar{z}} & c_{y\bar{z}} & 0 & 0 & 0\\
0 & 0 & c_{x\bar{y}} & -c_{xz} & 0 & -c_{xz} & c_{y\bar{z}} & 0 & 0 & 0 & 0 & 0\\
c_{x\bar{y}} & -c_{x\bar{y}} & 0 & 0 & c_{xz} & 0 & -c_{y\bar{z}} & 0 & 0 & 0 & 0 & 0
\end{array}\right)
\end{eqnarray}
Note that here we use simplified notations defined as follows:
$c_{\alpha\beta}=\cos(k_\alpha+k_\beta)$, $c_{\alpha{\bar \beta}}=\cos(k_\alpha-k_\beta)$, ${\tilde c}_{\alpha\beta}=\cos(2(k_\alpha+k_\beta))$, and  $s_{\alpha\beta}=i\sin(k_\alpha+k_\beta)$.

\section{Finite-difference method for integrating LLG equation} 
\label{a1}

In this appendix, we discuss a finite difference scheme for numerical
integration of the Landau-Lifshitz-Gilbert(LLG) equation by adopting
the method introduced by Serpico {\it et
al}\cite{serpico01}. The LLG equation can be written
in the following form:

\begin{equation}\label{eq:a1_LLG}
\frac{\partial{\bf S}_{i}}{\partial t}={\bf S}_{i}\times\left({\bf H}_{i}+\frac{\alpha_{G}}{S}\frac{\partial{\bf S}_{i}}{\partial t}\right),
\end{equation}
Note that the effective exchange field ${\bf H}_{i}$ is defined in
the Eq. (\ref{eq:H-eff}).

Then the following approximations are applied on the midpoint between
time step $n$ and $n+1$ :

\begin{equation}\label{eq:midpoint_approx1}
\frac{\partial{\bf S}_{i}}{\partial t}(t_{n+1/2})=\frac{{\bf S}_{i}(t_{n+1})-{\bf S}_{i}(t_{n})}{\Delta t}+O(\Delta t^{2})
\end{equation}

\begin{equation}\label{eq:midpoint_approx2}
{\bf S}_{i}(t_{n+1/2})=\frac{{\bf S}_{i}(t_{n+1})+{\bf S}_{i}(t_{n})}{2}+O(\Delta t^{2})
\end{equation}

where $\Delta t=t_{n+1}-t_{n}$ and $t_{n+1/2}$ denotes midpoint
between $t_{n}$ and $t_{n+1}$.

We now express ${\bf H}_{i}(t_{n+1/2})$ in terms of ${\bf H}_{i}(t_{n})$
and ${\bf H}_{i}(t_{n-1})$ by applying the midpoint approximation (\ref{eq:midpoint_approx2}) twice
:

\begin{equation}\label{eq:midpoint_approx3}
{\bf H}_{i}(t_{n+1/2})=\frac{3}{2}{\bf H}_{i}(t_{n})-\frac{1}{2}{\bf H}_{i}(t_{n-1})
\end{equation}

By substituting Eq. (\ref{eq:midpoint_approx1}),(\ref{eq:midpoint_approx2}), and (\ref{eq:midpoint_approx3}) into Eq. (\ref{eq:a1_LLG}) we get:

\begin{equation}
\frac{{\bf S}_{i}(t_{n+1})-{\bf S}_{i}(t_{n})}{\Delta t}=\left[\frac{{\bf S}_{i}(t_{n+1})+{\bf S}_{i}(t_{n})}{2}\right]\times\left[\frac{3}{2}{\bf H}_{i}(t_{n})-\frac{1}{2}{\bf H}_{i}(t_{n-1})+\frac{\alpha_{G}}{S}\frac{{\bf S}_{i}(t_{n+1})-{\bf S}_{i}(t_{n})}{\Delta t}\right]
\end{equation}

which yields the expression for ${\bf S}_{i}(t_{n+1})$ in terms of
previous time step spin components ${\bf S}_{i}(t_{n})$ and ${\bf S}_{i}(t_{n-1})$:

\begin{equation}
{\bf S}_{i}(t_{n+1})-{\bf S}_{i}(t_{n+1})\times\left[\left(\frac{3}{4}{\bf H}_{i}(t_{n})-\frac{1}{4}{\bf H}_{i}(t_{n-1})\right)\Delta t-\frac{\alpha}{S}{\bf S}_{i}(t_{n})\right]
\end{equation}

\begin{equation}
={\bf S}_{i}(t_{n})+{\bf S}_{i}(t_{n})\times\left(\frac{3}{4}{\bf H}_{i}(t_{n})-\frac{1}{4}{\bf H}_{i}(t_{n-1})\right)\Delta t
\end{equation}

Then the spin configuration ${\bf S}_{i}\left(t_{n+1}\right)$ is
obtained by solving the matrix equation:

\begin{equation}
{\bf S}_{i}(t_{n+1})={\bf B}_{i}(t_{n},t_{n-1})^{-1}{\bf A}_{i}\left(t_{n},t_{n-1}\right){\bf S}_{i}\left(t_{n}\right),
\end{equation}
where ${\bf A}_{i}(t_{n},t_{n-1})$ and ${\bf B}_{i}\left(t_{n},t_{n-1}\right)$
are both $3\times3$ matrix defined as follows :

\begin{equation}
{\bf A}_{i}\left(t_{n},t_{n-1}\right)=\left(\begin{array}{ccc}
1 & -\mathcal{K}_{i}^{z} & \mathcal{K}_{i}^{y}\\
\mathcal{K}_{i}^{z} & 1 & -\mathcal{K}_{i}^{x}\\
-\mathcal{K}_{i}^{y} & \mathcal{K}_{i}^{x} & 1
\end{array}\right),
\end{equation}

\begin{equation}
{\bf B}_{i}(t_{n},t_{n-1})=\left(\begin{array}{ccc}
1 & \mathcal{F}_{i}^{z} & -\mathcal{F}_{i}^{y}\\
-\mathcal{F}_{i}^{z} & 1 & \mathcal{F}_{i}^{x}\\
\mathcal{F}_{i}^{y} & -\mathcal{F}_{i}^{x} & 1
\end{array}\right),
\end{equation}
where $\hat{{\bf \mathcal{K}}}_{i}=\left[\frac{1}{4}{\bf H}_{i}(t_{n-1})-\frac{3}{4}{\bf H}_{i}(t_{n})\right]\Delta t$
and $\hat{{\bf \mathcal{F}}}_{i}=\hat{{\bf \mathcal{K}}}_{i}+\frac{\alpha_{G}}{S}{\bf S}_{i}(t_{n})$.

\section{Torques in the linearized LLG equation}\label{a2}

The torques in the linearized time-dependent LLG equation~(\ref{eq:LLG-sigma}) are calculated in this Appendix.
First we choose the following parametrization for the local basis:
\begin{eqnarray}\label{LocalCoordinates}
\begin{array}{l}
	{\bf \hat e}^x_{i}=\cos\theta_{i}\cos\phi_{i} {\bf \hat a}+\cos\theta_{i}
	\sin\phi_{i} {\bf \hat b}-\sin\theta_{i}{\bf \hat c}~,\\
	{\bf \hat e}^y_{i}=-\sin\phi_{i}{\bf \hat a}+\cos\phi_{i}{\bf \hat b}~,\\
	{\bf \hat n}_{i}=\sin\theta_{i}\cos\phi_{i}{\bf \hat a}+
	\sin\theta_{i}\sin\phi_{i}{\bf \hat b}+\cos\theta_{i}{\bf \hat c}~,
\end{array}
\end{eqnarray}
where $\theta_{i}$ and $\phi_{i}$ are polar and azimuthal angles that determine the orientation
of spin at site $i$. Introducing the following notations for simplicity:
$c_i=\cos\theta_{i}$, $s_i=\sin\theta_{i}$, ${\tilde c}_i=\cos\phi_{i}$, ${\tilde s}_i=\sin\phi_{i}$.
the linearized torques projected onto the localized basis are:

\begin{eqnarray}
	\tau_{i}^{x} &=& \sum_{j}\Bigl[D_{ij}^{z}c_{j}{\tilde c}_{i-j} + \left(D_{ij}^{x}{\tilde c}_{i}+
	D_{ij}^{y}{\tilde s}_{i}\right) s_{j}+J_{ij}c_{j}{\tilde s}_{i-j}\Bigr]\,\sigma_{j}^{x}
	+ \sum_{j}\Bigl[D_{ij}^{z}{\tilde s}_{i-j}-J_{ij}{\tilde c}_{i-j}\Bigr]\sigma_{j}^{y}
	\nonumber \\
	&+& \sum_{j} \Bigl[\left(D_{ij}^{x}{\tilde s}_{i}-D_{ij}^{y}{\tilde c}_{i}\right) s_{i} c_{j}
	D_{ij}^{z}s_{i}s_{j}{\tilde s}_{i-j} + \left(D_{ij}^{y}{\tilde c}_{j}- D_{ij}^{x}{\tilde s}_{j}\right)\,c_{i}s_{j}
	+J_{ij}\left(c_{i}c_{j}+s_{i}s_{j}{\tilde c}_{i-j}\right)\Bigr] \sigma_{i}^{y} ,
	\\	\nonumber \\ \nonumber \\
	\tau_{i}^{y} &=& \sum_{j}\Bigl[\left(D_{ij}^{y}{\tilde c}_{i}-D_{ij}^{x}{\tilde s}_{i}\right)s_{i}c_{j}
	+D_{ij}^{z}s_{i}s_{j}{\tilde s}_{i-j}
	+\left(D_{ij}^{x}{\tilde s}_{j}-D_{ij}^{y}{\tilde c}_{j}\right)c_{i}s_{j}-J_{ij}\left(c_{i}c_{j}+s_{i}s_{j}{\tilde c}_{i-j}\right)	\Bigr]\sigma_{i}^{x}
	\nonumber \\
	&+& \sum_{j}\Bigl[\left(D_{ij}^{y}{\tilde c}_{i}-D_{ij}^{x}{\tilde s}_{i}\right)
	c_{i}s_{j}-\left(D_{ij}^{y}s_{i}+D_{ij}^{z}c_{i}{\tilde s}_{i}\right)c_{j}{\tilde c}_{j}
	+\left(D_{ij}^{z}c_{i}{\tilde c}_{i}+D_{ij}^{x}s_{i}\right)c_{j}{\tilde s}_{j}+
	J_{ij}\left(c_{i}c_{j}{\tilde c}_{i-j}+s_{i}s_{j}\right)\Bigr]\,\sigma_{j}^{x}
	\nonumber\\
	&+& \sum_{j} \Bigl[D_{ij}^{z}c_{i}{\tilde c}_{i-j}+\left(D_{ij}^{x}{\tilde c}_{j}+
	D_{ij}^{y}{\tilde s}_{j}\right)s_{i}+J_{ij}c_{i}{\tilde s}_{i-j}\Bigr]\sigma_{j}^{y}
\end{eqnarray}

\end{widetext}

\end{document}